\documentclass[conference]{IEEEtran}
\usepackage{graphics}
\usepackage{graphicx}
\usepackage{epsfig,amssymb}
\usepackage{epstopdf}
\usepackage{amsmath}
\usepackage{breqn}
\usepackage{times}
\usepackage{mathrsfs}
\usepackage{array}
\usepackage{amsmath, latexsym, amsfonts, amssymb}
\usepackage[mathscr]{eucal}
\usepackage{array, tabularx}
\usepackage[table]{xcolor}
\usepackage{multirow}
\usepackage{epsfig}

\usepackage{cite}
\usepackage[numbers,sort & compress]{natbib}

\usepackage{tikz}
\usepackage{url}
\usepackage{mathtools}
\usepackage{psfragx}
\usepackage{xcolor}
\usepackage{bbm}
\usepackage{authblk}
\usepackage{balance}

\newcommand{\sqparen}[1]{\left[#1\right]}
\newcommand{\brparen}[1]{\left\{#1\right\}}

\renewcommand{\vec}[1]{\ensuremath{\boldsymbol{#1}}} %Re-define \vec command to generate vectors in bold

\IEEEoverridecommandlockouts

\begin{document}

\title{RF-based Direction Finding of UAVs Using DNN \bigskip}

%\title{RF-Based Direction Finding of UAVs Using Deep Neural Networks}

%\author{
%	\IEEEauthorblockN{Samith Abeywickrama}
%	\IEEEauthorblockA{Singapore University of Technology \\and Design, Singapore.\\
%		Email: abeywickrama$ \_ $samith@sutd.edu.sg}\and
%	
%	\IEEEauthorblockN{Lei Liu}
%	\IEEEauthorblockA{Singapore University of Technology \\and Design, Singapore.\\
%		Email: lei$ \_ $liu@sutd.edu.sg}\and	
%	
%	
%	\IEEEauthorblockN{Chau Yuen}
%	\IEEEauthorblockA{Singapore University of Technology \\and Design, Singapore.\\
%		Email: yuenchau@sutd.edu.sg}
%
%	 
%	\vspace{-3cm}
%}

%\author[$ $]{Samith Abeywickrama}
%\author[$  $]{Lahiru Jayasinghe}
%\author[$   $]{Hua Fu}
%\author[$  $]{Chau Yuen}
%\affil[$  $]{Singapore University of Technology and Design,  Singapore}
%%\affil[$ \ast $]{National University of Singapore, Singapore  }
%
%\affil[$  $ ]{Email: \textit { tharindu@mymail.sutd.edu.sg}, \textit { \{aruna$ \_ $jayasinghe,hua$ \_ $ fu,yuenchau\}@sutd.edu.sg}  
%	%\thanks{The first two authors contributed equally.}   
%}

\author[$\ast \dag $]{Samith Abeywickrama}
\author[$ \dag $]{Lahiru Jayasinghe}
\author[$ \dag  $]{Hua Fu}
\author[$ \dag $]{Subashini Nissanka}
\author[$ \dag $]{Chau Yuen}
\affil[$ \dag $]{Singapore University of Technology and Design,  Singapore}
\affil[$ \ast $]{National University of Singapore, Singapore  }

\affil[$  $ ]{Email: \textit { samith@u.nus.edu}, \textit { \{aruna$ \_ $jayasinghe,hua$ \_ $ fu,yuenchau\}@sutd.edu.sg}  
	\thanks{The support of NSFC 61750110529 and SUTD-MIT International Design
		Center is gratefully acknowledged.
		The first two authors contributed equally.}   
	}

%\author[$ \star \ddag $]{Samith Abeywickrama}
%\author[$  $]{Lahiru Jayasinghe}
%\author[$  $]{Hua Fu}
%\author[$  $]{Chau Yuen}
%\affil[$  $]{Singapore University of Technology and Design,  Singapore}
%%\affil[$ \ddag $]{Nanyang Technological University, Singapore}
%\affil[$  $ ]{Email: \textit { tharindu@mymail.sutd.edu.sg}, \textit { \{aruna$ \_ $jayasinghe,hua$ \_ $ fu,yuenchau\}@sutd.edu.sg}}
%%\affil[$ \ddag $ ]{Email: \textit { \{chiyuhao1990\}@163.com} 
%%\thanks{This work is supported by A*STAR (Agency for Science, Technology and Research) SERC project, under the grant no. 1420200043.}

\date{}
\bibliographystyle{ieeetr}
\maketitle

\begin{abstract} 
	This paper presents a sparse denoising autoencoder (SDAE)-based deep neural network (DNN) for the direction finding (DF) of small unmanned aerial vehicles (UAVs). It is motivated by the practical challenges associated with classical DF algorithms such as MUSIC and ESPRIT.  The proposed DF scheme is practical and low-complex in the sense that a phase synchronization mechanism, an antenna   calibration mechanism, and the analytical model of the antenna radiation pattern  are not essential. Also, the proposed DF method can be implemented using a single-channel RF receiver.	The paper validates the proposed method experimentally as well.
\end{abstract}

\bigskip
\bigskip

\begin{IEEEkeywords}
	Drone surveillance, direction finding, UAV tracking
\end{IEEEkeywords}

\section{Introduction}
\label{sec:intro}

The use of civilian drones has increased dramatically in recent years. Likewise, drones are fast gaining popularity around the world \cite{yingxiang_iccs,wu2018joint,wu2,wu3,wu4}. However, drone use in a problematic manner has stirred public concerns. For example, in January 2015 a drone crashed at the White House \cite{whouse}, raising concerns about security risks to the government building; in March 2016 a Lufthansa jet came within 200 feet of colliding with a drone near Los Angeles International Airport \cite{jet}; and drones have been accused of being used to violate the privacy and even carry criminal activities \cite{couple}. These events give ample self-evident examples that developing a surveillance system for suspect drones is of paramount importance. 

The authors of \cite{MagDrone} sought to detect a drone using Radio Frequency (RF) as it can work day and night and at all weather conditions. Most of the commercial drones communicate frequently with their controllers, and the downlink, \textit{i.e.}, video signal and telemetry signals (flight speed, position, altitude, and battery level),  between the drone and its controller is always present. To this end, this paper presents a drone surveillance system, by eavesdropping on the communication between a drone and its ground controller. The system can  estimate drone's direction (or bearing) by processing the data transmitted from the drone to its controller using a single channel wireless receiver. Therefore, no dedicated transmitter is required at the surveillance system.

RF based direction finding (DF) techniques have been well studied, and the classical high-resolution techniques such as MUSIC \cite{music} and ESPRIT \cite{isprt} are considered to be the most popular algorithms. However, MUSIC  and ESPRIT are inherently multi-channel techniques because  those algorithms require a snapshot observation.  This means, the base-band data from all antenna elements should be extracted simultaneously so that a data correlation matrix can be formulated. Therefore, multiple channels should be coherent. However, in most receivers, the digital down converter (DDC) chain uses a coordinate rotation digital computer (CORDIC), which has a random start-up position on power up. The CORDIC therefore creates a random phase each time when the channels of the receiver are initialized, but remains constant throughout the operation \cite{cordic}. Therefore, calibrating  this start-up phase values of each RF channel becomes necessary to realize a coherent multi-channel receiver.  Clearly, this  increases the hardware complexity and the power consumption.

Most of the civilian drones use WiFi-like OFDM for their communication. They are usually unknown, wideband,  and transmitted in burst-mode. Such signal characteristics  pose challenges with classical DF techniques. However, if only the signal power measurements are utilized, performing DF  is practically feasible even with such signals \cite{FuHua_iccs,power2,2017arXiv170609690P,power1}. In \cite{power1}, signal power measurements that are obtained from a switched beam antenna array are utilized to estimate the direction of a WiFi transmitter. As the actual   radiation pattern of the antenna is vital for these methods, still it bounds with some practical challenges. Therefore, we propose a practical and a low-complex drone DF method in this paper, and our contributions can be summarized as fallows.

To the best of authors' knowledge there has not been any other method that involves deep neural network  in the context of drone DF. We focus on a system which comprises a directional antenna array having $ N $  antennas, and a single channel receiver. By processing the signals that are transmitted from the drone to its ground controller, the single channel  receiver measures the received signal power  at the each antenna using a RF switching mechanism. Then, the obtained  power values are fed to the proposed sparse denoising autoencoder (SDAE)-based deep neural network (DNN). More precisely, the first hidden layer of the network extracts a  robust sparse representation of  the received power values. Then, the rest of the network utilizes this sparse representation  to classify  the direction of the drone signal. It should be noted that a phase
synchronization mechanism, an antenna gain calibration mechanism, and the analytical model of the antenna radiation pattern are not essential for this single channel implementation. The paper  validates the proposed method experimentally through a software defined radio (SDR) implementation in conjunction with TensorFlow \cite{tenserflow}. Furthermore, such an experimental validation for  drone DF is not common in the literature, and can be highlighted as another  contribution of this paper.

The paper organization is as follows. The system model is presented in Section \ref{problem}. Section \ref{archi} discusses the proposed deep architecture. Then, in Section \ref{exper}, we validate the proposed method using  experimental results. Section \ref{conc} concludes the paper.

To promote reproducible research, the codes for generating most of the results in the paper are made
available on the website: https://github.com/LahiruJayasinghe/DeepDOA.

%
%%Most of the civilian drones operate in the Industrial, Scientific and Medical (ISM) band, and they are using Wifi-like OFDM. In practice, statistical parameters (mean, variance, autocorrelation, etc.) of most manmade signals are assumed to be varying in time with single or multiple periodicities [16] due to periodic keying of the amplitude, repeating spreading codes, repeating preambles, etc. Also, most of the drones usually communicate with their controllers frequently around 30 times per second to update its status as well as to receive the commands from controller [14].
%%Therefore, drone signals can be modelled as cyclostationary processes due to its inherent periodicities. In this paper, we employ cyclostationary feature analysis for the detection of drones.
%

%This is not the case for the single-channel implementation since only one element data sample is extracted at a time.
%
%For DoA estimation,  
%
%multi-beam antenna system
%
%After the detection of drone's presence, we employ

% Therefore, assuming that we have a prior knowledge about the common nature of periodicities of drone communications, calculating the autocorrelation of the frequency band of interest allows us to identify drone communications by analyzing underlying periodicities.

\section{System Model}
\label{problem}

\begin{figure}[t] \vspace{0.5cm}
	\centering {\includegraphics[scale=0.87]{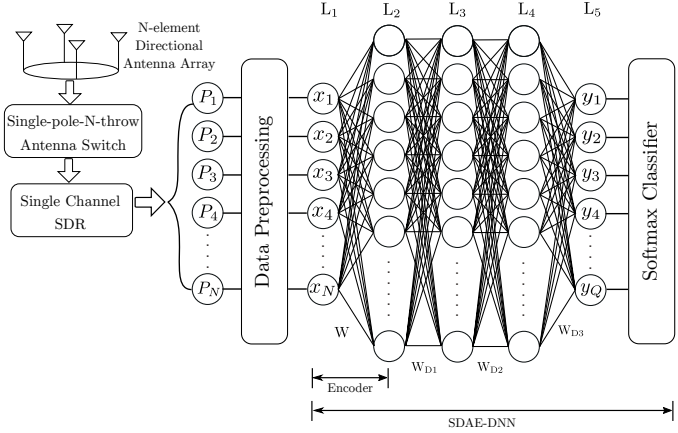}} 
	\caption{The System Model.}	 
	\label{sm}
\end{figure}

We consider a system which consists of a  single channel receiver, and a circular antenna array equipped with $ N $ direction antennas, see Fig. \ref{sm}. The antenna array is connected to the receiver using a non-reflective Single-Pole-N-Throw (SPNT) RF switch. The  switching period is $ T_s $. Suppose that a far-field drone signal  impinges on the antenna array with azimuth angle $ \theta \in [0 \enspace 2\pi) $. The received signal at the $n-$th antenna element can be given as 
\begin{equation}
r_n(k) = a_n(\theta) s(k) + n_n(k),
\end{equation}
where  $ k $ is the sample index, $ a_n(\theta) $ is the $ n-$th antenna response vector for the azimuth angle $ \theta $, $ s(k) $ is the drone transmitted signal as it arrives at the antenna array, $ n_n(k) $ is circularly symmetric, independent and identically distributed,
complex additive white Gaussian noise (AWGN) with zero mean and variance $ \sigma^2 $, and $ n \in \{1, \cdots,N \} $. Here, $  a_n(\theta) $ follows the form of 
\begin{equation}
a_n(\theta) = G_n(\theta) e^{j \frac{2\pi}{\lambda}\beta_n(\theta)},
\end{equation}
where $ G_n(\theta) $ is the real numbered antenna gain for the azimuth angle $ \theta $, $ \lambda $ is the signal wavelength, and
\begin{equation}
\beta_n(\theta) = d \cos \Bigg[ \frac{2\pi(n-1)}{N} - \theta \Bigg]   ,
\end{equation}
where  $ d $ is the radius of the circular antenna array \cite{gain}. Since we consider  a  practical DF method in this paper, $ s(k) $ and $ a_n(\theta) $ are assumed to be unknown. Therefore, our objective is to recover the  azimuth angle $ \theta $, while the parameters $ s(k) $ and $ a_n(\theta) $ are unknown. 

%Note that $ \theta $, $ s(k) $, and $ a_n(\theta) $ are assumed to be unknown in practice. The aim of the drone direction finding is to recover the unknown azimuth angle $ \theta $.

We focus on a power measurements based approach. To this end,  the ensemble averaged received signal power at the $n-$th antenna element can be given as
\begin{alignat}{5}  \label{powe}
P_n  &= \mathrm{E} \Big[ |r_n(k)|^2 \Big]   \nonumber\\
&= \mathrm{E} \Big[ \Big( a_n(\theta) s(k) + n_n(k) \Big) \Big( a_n(\theta) s(k) + n_n(k) \Big)^{\ast} \Big]      \nonumber\\
&= |a_n(\theta)|^2 \mathrm{E} \Big[|s(k)|^2 \Big]   + \mathrm{E} \Big[ |n_n(k)|^2 \Big]  \nonumber\\
&= G_n^2(\theta) P_s + \sigma^2, 
\end{alignat} 
%\begin{alignat}{5}  \label{powe}
%P_n  &= \frac{1 }{K} \sum_{k=1}^{K}  |r_n(k)|^2    \nonumber\\
%&= \frac{1 }{K} \sum_{k=1}^{K} \Big[ \Big( a_n(\theta) s(k) + n_n(k) \Big) \Big( a_n(\theta) s(k) + n_n(k) \Big)^{\ast} \Big]      \nonumber\\
%&= |a_n(\theta)|^2 \frac{1 }{K}  \sum_{k=1}^{K}  |s(k)|^2  + \frac{1 }{K}  \sum_{k=1}^{K}  |n_n(k)|^2   \nonumber\\
%&= G_n^2(\theta) P_s + \frac{1 }{K}  \sum_{k=1}^{K}  |n_n(k)|^2 \nonumber\\
%&= G_n^2(\theta) P_s + \sigma^2, 
%\end{alignat} 
where  $$ G_n^2(\theta) = |a_n(\theta)|^2,  $$ $$ P_s = \mathrm{E} \Big[|s(k)|^2 \Big], $$  $$ \sigma^2= \mathrm{E} \Big[ |n_n(k)|^2 \Big],$$ and $ \mathrm{E} [\cdot]  $ denotes the expectation operator. Here, \eqref{powe} follows  from the fact that  $ s(k) $ and $ n_n(k) $ are  independent and uncorrelated, \textit{i.e.}, $$ \mathrm{E} [s(k)n_n^{\ast}(n)] = \mathrm{E} [s^{\ast}(k)n_n(n)] =0. $$ It can be observed that the received power values at the antenna elements $ n_i $ and $ n_j $, where $ i,j \in \{1,\cdots,N \} $ and $ i \neq j $,  are not identical, since $ G_i^2(\theta) \neq G_j^2(\theta) $.
%(directional antenna elements are pointing toward different directions). 
We have this property thanks to the gain variation of the directional antenna array in $ [0 \enspace 2\pi) $.  Therefore, it is desirable to have an underlying relationship (or a pattern) between $\brparen{P_n}_{n=1}^N$  and $ \theta $. %Furthermore, the incoming drone signal with direction $\theta$ occupies a certain isolated point in the angle domain of  $ [0 \enspace 2\pi) $. Therefore, $ \theta $ is sparse in the spatial domain. Our proposed method  is inspired  by above two properties, and it can be summarized as follows.  

The proposed method is as follows.
The receiver sequentially activates one antenna element at a time using the SPNT RF switch, and measures the corresponding received power value. During the activation of $ n-$th antenna, $ P_n $ is measured, where $ n \in \{1, \cdots,N \} $.  A single switching cycle is equivalent to $ N $ activations, starting from  the first antenna to the $ N-$th antenna. Let  $ \mathrm{{\textbf{p}}}  = [P_1,\cdots,P_N]^\top  $ denote the power measurements corresponding  to a single switching cycle. As it is depicted in Fig. \ref{sm}, $ \vec{x} = [x_1,\cdots,x_N]^\top  $ is obtained during the preprocessing stage, where 
\begin{equation}
x_n = \frac{P_n}{\sum_{i=1}^{N} P_i}.
\end{equation}
This means,  $ x_n $  is the ratio between  $ P_n $ and the summation of all  power values within the same switching cycle. 
%Next, $ \vec{x} $ is fed to the input  of the deep network.
In the next section, we discuss how  the proposed network  recovers $ \theta $ from  $ \vec{x} $.

%Starting from a signal power measurements vector $ \vec p = \sqparen{p_{1},\ldots, p_{N}}^\top $ as described in the previous section, we compute $\vec{x}= \sqparen{x_{1},\ldots, x_{N}}^\top $.

\section{SDAE-DNN Architecture}
\label{archi}

\begin{figure}[t] \vspace{0.3cm}
	\centering {\includegraphics[scale=0.8]{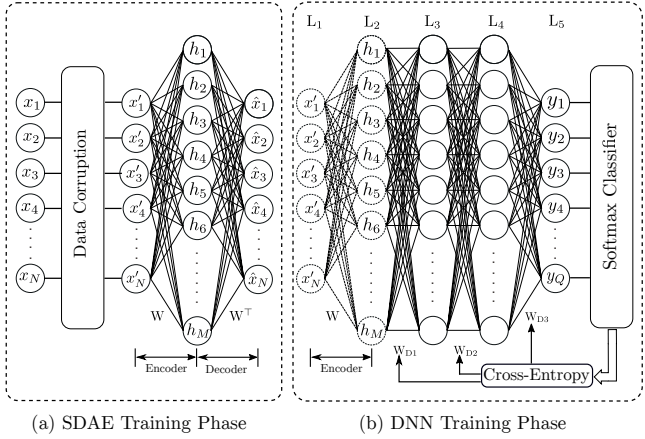}} 
	\caption{Training Phases.}	 
	\label{tr}
\end{figure}

The proposed deep architecture comprises a trained SDAE  and a trained DNN, followed by a fully-connected softmax classifier layer, see Fig. \ref{sm}.  During the training phase of the SDAE (Fig. \ref{tr}-(a)), the preprocessed received power values $\brparen{x_n}_{n=1}^N$ are assigned to  the input units. Therefore, the number of neurons in the input layer is equal to the number of elements $ N $ in the directional antenna array. Then,  the values of the  hidden layer units are calculated as
\begin{equation}\label{equ1}
\vec h = f(\mathrm {\mathbf W} f_c(\vec x) + \vec {b_e}),
\end{equation} 
and output layer values are calculated as
\begin{equation}\label{equ2}
\vec {\hat x} = f(\mathrm {\mathbf W}^\top \vec h + \vec {b_d}), 
\end{equation}
where $ f(\cdot) $  is non-linear  activation function  that operates element-wise on its argument, $\mathrm {\mathbf W} \in \mathbb R^{M\times N} $ denotes the encoder  weight matrix, $ \vec {b_e} = \sqparen{b_{e_1},\ldots, b_{e_M}}^\top $ and $ \vec {b_d} = \sqparen{b_{d_1},\ldots, b_{d_N}}^\top $ denote the bias vectors, and  $ f_c(\cdot) $ is a stochastic corrupter which adds noise according to some noise model to its input, \textit{i.e.}, $ \vec { x'} = f_c(\vec x) $, where $ \vec { x'} = \sqparen{ x^'_1,\ldots, x^'_N}^\top $.  In \eqref{equ1}, $ f_c $ is non-deterministic, since it corrupts the same set of received power values $\brparen{x_n}_{n=1}^N$ in different ways every time $\brparen{x_n}_{n=1}^N$ is passed through it. $\mathrm {\mathbf W^\top}  $ is the decoder weight matrix, which ensures that the output layer reconstructs the input as precisely as possible ($ \mathrm {\mathbf W}^{\top} $ is the matrix transpose of $ \mathrm {\mathbf W} $).  Here, we particularly target on reconstructing the input received power values at the output layer of the SDAE. 
%Therefore, data labelling is not an essential task in this  phase, and hence, the learning strategy is unsupervised.  

To this end, the parameters of  SDAE ($\mathrm {\mathbf W}  $, $ \vec {b_e} $, and $ \vec {b_d}) $ are optimized such that the  reconstruction error is minimized, while it  subjecting  to a sparsity constraint. This sparsity constraint  encourages the sparse activation of the hidden layer units. Therefore, the  cost function  can be given as
\begin{equation}
L(\mathrm {\mathbf W}, \vec {b_e}, \vec {b_d}) = \sum_{i=1}^{T} ( \vec {\hat x_i} - \vec {x_i})^2 + \beta \sum_{m=1}^{M} \mathrm{KL} (\rho || \rho_m), \label{los}
\end{equation}
where $ T $ is the size of the training data set, $\beta $ is a hyper parameter\footnote{ $ \beta $ operates as the trade-off parameter between the squared error and $ \mathrm{KL} (\rho || \rho_m) $, and its value can be empirically decided during the training process.}, $ \rho $ is the sparsity parameter, $$ \rho_m  = \frac{1}{T} \sum_{i=1}^{T} h_m(\vec {x_i}) $$ is the average activation level of the $ m $-th hidden unit 
%over the training data set
where $ h_m(\vec {x_i}) $ denotes the activation of the $ m- $th  unit for the input $ \vec {x_i} $, and 

\begin{equation}
\mathrm{KL} (\rho || \rho_m) = \rho \log  \Big(\frac{\rho}{\rho_m}\Big)+(1-\rho)\log\Big(\frac{1-\rho}{1-\rho_m}\Big) \label{kl}
\end{equation}
is the Kullback-Leibler (KL) divergence \cite{kl}. From \eqref{kl}, it can be observed that $ \mathrm{KL} (\rho || \rho_m) = 0 $, if $ \rho_m = \rho $, and otherwise it increases monotonically as $ \rho_m $ diverges from $ \rho $. Typically, $ \rho $ is a very small value close to zero. 
Therefore, when the cost function \eqref{los} is minimized, the parameter $ \rho $   enforces   $\brparen{\rho_m}_{m=1}^M$  to be close to zero, while the  dominant neurons that represent specific features  stay non-zero. Now, the decoder is discarded, and the trained encoder is connected to the DNN as a fully connected  layer.

%will be chosen empirically, and it

Next, the  DNN training phase   is commenced. As Fig. \ref{tr}-(b)  depicts, DNN comprises three fully connected hidden layers, \textit{i.e.}, $ \mathrm L_2 $, $ \mathrm L_3 $, and $ \mathrm L_4 $, and a softmax layer \cite{softmax} for the task of classification. Since $ \mathrm L_2 $ is an element  of the trained encoder,  it uses the same activation function $ f $. Hidden layers $ \mathrm L_3 $ and $ \mathrm L_4 $  use Rectified liner Unit  (ReLU) \cite{7280578,toilet,choraya2} as their activation function. Again, noise corrupted received power values $\brparen{x^'_n}_{n=1}^N$ are   the training inputs.  Now, data need to be labelled into $ Q $ classes due to the use of softmax classifier, where the label is the direction of the drone signal coming from. Therefore, the learning strategy is supervised in this training phase. Since,  $ \mathrm {\mathbf { W } }  $ is the pre-trained encoder weight matrix, it will not be optimized again. Therefore,  only the  weight matrices $ \mathrm {\mathbf { W_{\mathrm{D1}} } }  $, $ \mathrm {\mathbf { W_{\mathrm{D2}} } }  $, and $ \mathrm {\mathbf { W_{\mathrm{D3}} } }  $ are optimized during this training phase.

\textit{\underline{Remark 1:}} It should be noted that the incoming drone signal with direction $\theta$ occupies a certain isolated point in the angle domain of  $ [0 \enspace 2\pi) $. Therefore, $ \theta $ is sparse in the spatial domain, and this sparsity can be exploited to estimate $ \theta $. Here, we use this sparse property, and it can be summarized as follows. In the cost function \eqref{los}, the squared error is calculated between the non-corrupted power values and the reconstructed power values, while the noise-corrupted  power values are fed to the network. This cost function is  subject to a sparsity constraint as well.  Therefore, even when the system operates in a noisy environment, the first hidden layer (or $ \mathrm {\mathbf { W } }  $) of the network extracts a  robust sparse representation of the input power values. 
%Therefore, the first layer can be recognized as a noise reduction mechanism, and a feature extractor. 
Then, the rest of the network utilizes this sparse representation to classify (or estimate) $ \theta $. 

In the next section, we will validate our proposed method using experimental results.

\section{Experimental Validation} \label{exper}

Our experimental setup comprises  a SDR (USRP B210), and a four element sector antenna, which is  a variant of the antenna implemented in \cite{antenna}.  We use only a single RF receiving channel of the SDR. Therefore, the SDR is connected to the antenna using a non-reflective Single-Pole-4-Throw (SP4T) RF switch. DJI Phantom 3 is considered as the target drone throughout the experiment. The drone downlink channels  occupy the bandwidth from 2.401 GHz to 2.481 GHz, each has 10 MHz bandwidth OFDM signal. This OFDM signal transmitted by the drone provides the main source to perform the DF task. 

\begin{table}[]
	\centering
	\caption{Confusion matrix when the proposed network is used.} 
	\label{cm1}
	\scalebox{0.95}{
		\begin{tabular}{c|c|c|c|c|c|c|c|c|}
			\cline{2-9}
			& \multirow{2}{*}{$ \enspace 0^0 $} & \multirow{2}{*}{$ 45^0 $} & \multirow{2}{*}{$ 90^0 $} & \multirow{2}{*}{$ 135^0 $} & \multirow{2}{*}{$ 180^0 $} & \multirow{2}{*}{$ 225^0 $} & \multirow{2}{*}{$ 270^0 $} & \multirow{2}{*}{$ 315^0 $} \\ [2ex] \hline
			\multicolumn{1}{|c|}{\multirow{2}{*}{$ 0^0 $}} & \multirow{2}{*}{\textbf{95}} & \multirow{2}{*}{\textcolor{blue}{1}}  & \multirow{2}{*}{\textcolor{blue}{1}} & \multirow{2}{*}{\textcolor{gray}{0}}  & \multirow{2}{*}{\textcolor{blue}{3}} & \multirow{2}{*}{\textcolor{gray}{0}} & \multirow{2}{*}{\textcolor{gray}{0}}  & \multirow{2}{*}{\textcolor{gray}{0}}  \\ [2ex]\hline
			\multicolumn{1}{|c|}{\multirow{2}{*}{$ 45^0 $}} & \multirow{2}{*}{\textcolor{blue}{1}} & \multirow{2}{*}{\textbf{97}} & \multirow{2}{*}{\textcolor{gray}{0}}  & \multirow{2}{*}{\textcolor{gray}{0}}  & \multirow{2}{*}{\textcolor{blue}{1}} & \multirow{2}{*}{\textcolor{blue}{1}} & \multirow{2}{*}{\textcolor{gray}{0}}  & \multirow{2}{*}{\textcolor{gray}{0}}  \\ [2ex]\hline
			\multicolumn{1}{|c|}{\multirow{2}{*}{$ 90^0 $}} & \multirow{2}{*}{\textcolor{blue}{1}} & \multirow{2}{*}{\textcolor{blue}{1}} & \multirow{2}{*}{\textbf{98}} & \multirow{2}{*}{\textcolor{gray}{0}}  & \multirow{2}{*}{\textcolor{gray}{0}}  & \multirow{2}{*}{\textcolor{gray}{0}}  & \multirow{2}{*}{\textcolor{gray}{0}}  & \multirow{2}{*}{\textcolor{gray}{0}}  \\ [2ex]\hline
			\multicolumn{1}{|c|}{\multirow{2}{*}{$ 135^0 $}} & \multirow{2}{*}{\textcolor{gray}{0}}  & \multirow{2}{*}{\textcolor{gray}{0}}  & \multirow{2}{*}{\textcolor{gray}{0}}  & \multirow{2}{*}{\textbf{100}} & \multirow{2}{*}{\textcolor{gray}{0}}  & \multirow{2}{*}{\textcolor{gray}{0}}  &\multirow{2}{*}{\textcolor{gray}{0}}   &\multirow{2}{*}{\textcolor{gray}{0}}   \\[2ex] \hline
			\multicolumn{1}{|c|}{\multirow{2}{*}{$ 180^0 $}} & \multirow{2}{*}{\textcolor{blue}{3}} & \multirow{2}{*}{\textcolor{blue}{3}} & \multirow{2}{*}{\textcolor{gray}{0}}  & \multirow{2}{*}{\textcolor{gray}{0}}  & \multirow{2}{*}{\textbf{92}} & \multirow{2}{*}{\textcolor{blue}{2}} & \multirow{2}{*}{\textcolor{gray}{0}}  & \multirow{2}{*}{\textcolor{gray}{0}}  \\ [2ex]\hline
			\multicolumn{1}{|c|}{\multirow{2}{*}{$ 225^0 $}} & \multirow{2}{*}{\textcolor{gray}{0}}  & \multirow{2}{*}{\textcolor{gray}{0}}  & \multirow{2}{*}{\textcolor{red}{4}} & \multirow{2}{*}{\textcolor{gray}{0}}  & \multirow{2}{*}{\textcolor{blue}{1}} & \multirow{2}{*}{\textbf{95}} &  \multirow{2}{*}{\textcolor{gray}{0}}  & \multirow{2}{*}{\textcolor{gray}{0}}  \\[2ex] \hline
			\multicolumn{1}{|c|}{\multirow{2}{*}{$ 270^0 $}} & \multirow{2}{*}{\textcolor{gray}{0}}  & \multirow{2}{*}{\textcolor{gray}{0}}  & \multirow{2}{*}{\textcolor{blue}{3}} & \multirow{2}{*}{\textcolor{gray}{0}}  & \multirow{2}{*}{\textcolor{gray}{0}}  & \multirow{2}{*}{\textcolor{blue}{1}} & \multirow{2}{*}{\textbf{95}} & \multirow{2}{*}{\textcolor{blue}{1}} \\ [2ex]\hline
			\multicolumn{1}{|c|}{\multirow{2}{*}{$ 315^0 $}} & \multirow{2}{*}{\textcolor{gray}{0}}  & \multirow{2}{*}{\textcolor{gray}{0}}  & \multirow{2}{*}{\textcolor{gray}{0}}  & \multirow{2}{*}{\textcolor{gray}{0}}  &\multirow{2}{*}{\textcolor{gray}{0}}   & \multirow{2}{*}{\textcolor{gray}{0}}  & \multirow{2}{*}{\textcolor{blue}{1}} & \multirow{2}{*}{\textbf{99}} \\[2ex] \hline
		\end{tabular}}
	\end{table}
	
\begin{figure}[t] \vspace{1cm}
			
			\begin{minipage}[b]{.5\linewidth}
				\centering
				\centerline{\includegraphics[width=4.2cm]{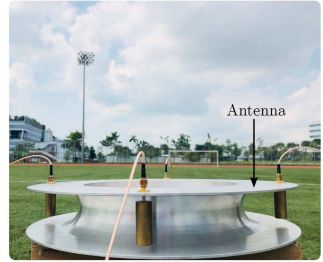}}
				%  \vspace{1.5cm}
				\centerline{(a) Training Field}\medskip
			\end{minipage}
			\hfill 
			\begin{minipage}[b]{0.48\linewidth}
				\centering
				\centerline{\includegraphics[width=3.5cm]{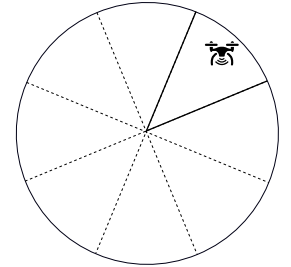}}
				%  \vspace{1.5cm}
				\centerline{(b) Direction Configuration}\medskip
			\end{minipage}
			\caption{The Training Field and The Direction Configuration.}
			\label{testing}
		\end{figure}

	%This experiment has been carried out under two phases. First, we trained the system, and then, we tested the system. 
	Fig. \ref{testing}-(a) represents the environment that we used for the training  data collection. This is a large ground with an open area. Also, there was negligible RF interference on the 2.401 GHz - 2.481 GHz range. To  simplify  the experiment, we virtually divided the area into eight octants, see Fig. \ref{testing}-(b). Each octant is considered as one direction  during the experiment. For example, the first octant is considered as $ 0 $ degrees direction, while the second octant is considered as $ 45 $ degrees direction, and so on. Therefore, when the drone is flying, its direction is indicated by its corresponding octant. 
	%Upon the completion of  data collection (received power values of the DJI Phantom 3 OFDM signal), data was normalized, and  corrupted  by adding  artificial noise.\footnote{Here we assume that the noise power follows the chi-squared distribution.} Then, we used them to train the  SDAE-DNN.  
	
	In the trained network, $ \mathrm L_5- $th layer has eight neurons (we have eight classes for the direction classification, or, $ Q=8 $) and $ \mathrm L_1- $th layer has four neurons (the antenna array has four elements, or, $ N=4 $). The hidden layers $ \mathrm L_2 $, $ \mathrm L_3 $, and $ \mathrm L_4 $ have $ 200 $,  $ 12 $, and $ 12 $ neurons, respectively. These values have been empirically decided during the training process.
	
	After the training phase, the evaluation is done in a different environment. Now, the frequency spectrum (2.401 GHz - 2.481 GHz) suffers from  WiFi and bluetooth interferences. To this end, two experiments have been carried out.  First, we evaluated the proposed deep architecture, and its confusion matrix is given in the Table I. Next, we considered a baseline method, where only a conventional DNN is implemented without the $ \mathrm L_2 $ layer (other layers have same number of nodes).  Its confusion matrix is given in the Table II.  Note that the confusion matries represent  the percentage (\%) values. It can be observed that the proposed deep architecture is certainly robust, and it outperforms the baseline method.
	
	Since our implementation does not use multiple RF channels and any information about the antenna radiation pattern,
	%Since we implement a  DF method without having the knowledge of  antenna radiation pattern, 
	it is not desirable to compare our results with conventional techniques. Therefore, we omit such simulation/experimental results. Further interesting experimental evaluations and insights will be presented in  future extensions of this work. 
	
	%with proposed SDAE-DNN and only a DNN to compare the performances and evaluate the noise reduction capability of the proposed architecture. As per the results shown in Table 1 and table 2, both methods have an aptitude of estimating the direction accurately. When we introduce a SDAE with DNN, 
	%the accuracy of the direction estimation has increased significantly. It means with the proposed architecture, in a noisy environment, the direction estimation accuracy is always greater than 92.

	\begin{table}[]
		\centering
		\caption{Confusion matrix when  only the DNN is used.} 
		\label{cm2}
		\scalebox{0.95}{
			\begin{tabular}{c|c|c|c|c|c|c|c|c|}
				\cline{2-9}
				& \multirow{2}{*}{$ \enspace 0^0 $} & \multirow{2}{*}{$ 45^0 $} & \multirow{2}{*}{$ 90^0 $} & \multirow{2}{*}{$ 135^0 $} & \multirow{2}{*}{$ 180^0 $} & \multirow{2}{*}{$ 225^0 $} & \multirow{2}{*}{$ 270^0 $} & \multirow{2}{*}{$ 315^0 $} \\ [2ex] \hline
				\multicolumn{1}{|c|}{\multirow{2}{*}{$ 0^0 $}} & \multirow{2}{*}{\textbf{94}} & \multirow{2}{*}{\textcolor{red}{5}}  & \multirow{2}{*}{\textcolor{blue}{1}} & \multirow{2}{*}{\textcolor{gray}{0}}  & \multirow{2}{*}{\textcolor{gray}{0}}  & \multirow{2}{*}{\textcolor{gray}{0}}  & \multirow{2}{*}{\textcolor{gray}{0}}  & \multirow{2}{*}{\textcolor{gray}{0}}  \\ [2ex]\hline
				\multicolumn{1}{|c|}{\multirow{2}{*}{$ 45^0 $}} & \multirow{2}{*}{\textcolor{red}{10}} & \multirow{2}{*}{\textbf{\textcolor{red}{88}}} & \multirow{2}{*}{\textcolor{blue}{1}} & \multirow{2}{*}{\textcolor{gray}{0}}  & \multirow{2}{*}{\textcolor{gray}{0}}  & \multirow{2}{*}{\textcolor{blue}{1}} & \multirow{2}{*}{\textcolor{gray}{0}}  & \multirow{2}{*}{\textcolor{gray}{0}}  \\ [2ex]\hline
				\multicolumn{1}{|c|}{\multirow{2}{*}{$ 90^0 $}} & \multirow{2}{*}{\textcolor{red}{4}} & \multirow{2}{*}{\textcolor{blue}{1}} & \multirow{2}{*}{\textbf{\textcolor{red}{88}}} & \multirow{2}{*}{\textcolor{blue}{1}} & \multirow{2}{*}{\textcolor{blue}{3}} & \multirow{2}{*}{\textcolor{gray}{0}}  &\multirow{2}{*}{\textcolor{gray}{0}}   & \multirow{2}{*}{\textcolor{blue}{3}} \\ [2ex]\hline
				\multicolumn{1}{|c|}{\multirow{2}{*}{$ 135^0 $}} & \multirow{2}{*}{\textcolor{gray}{0}}  & \multirow{2}{*}{\textcolor{red}{6}} & \multirow{2}{*}{\textcolor{blue}{1}} & \multirow{2}{*}{\textbf{93}} & \multirow{2}{*}{\textcolor{gray}{0}}  & \multirow{2}{*}{\textcolor{gray}{0}}  & \multirow{2}{*}{\textcolor{gray}{0}}  & \multirow{2}{*}{\textcolor{gray}{0}}  \\[2ex] \hline
				\multicolumn{1}{|c|}{\multirow{2}{*}{$ 180^0 $}} & \multirow{2}{*}{\textcolor{red}{30}} & \multirow{2}{*}{\textcolor{red}{4}} & \multirow{2}{*}{\textcolor{blue}{3}} & \multirow{2}{*}{\textcolor{gray}{0}}  & \multirow{2}{*}{\textbf{\textcolor{red}{60}}} & \multirow{2}{*}{\textcolor{blue}{2}} & \multirow{2}{*}{\textcolor{gray}{0}}  & \multirow{2}{*}{\textcolor{blue}{1}} \\ [2ex]\hline
				\multicolumn{1}{|c|}{\multirow{2}{*}{$ 225^0 $}} & \multirow{2}{*}{\textcolor{blue}{1}} & \multirow{2}{*}{\textcolor{gray}{0}}  & \multirow{2}{*}{\textcolor{red}{4}} & \multirow{2}{*}{\textcolor{gray}{0}}  & \multirow{2}{*}{\textcolor{blue}{2}} & \multirow{2}{*}{\textbf{91}} & \multirow{2}{*}{\textcolor{gray}{0}}  & \multirow{2}{*}{\textcolor{blue}{2}} \\[2ex] \hline
				\multicolumn{1}{|c|}{\multirow{2}{*}{$ 270^0 $}} & \multirow{2}{*}{\textcolor{gray}{0}}  & \multirow{2}{*}{\textcolor{gray}{0}}  & \multirow{2}{*}{\textcolor{blue}{1}} & \multirow{2}{*}{\textcolor{gray}{0}}  & \multirow{2}{*}{\textcolor{blue}{2}} & \multirow{2}{*}{\textcolor{blue}{1}} & \multirow{2}{*}{\textbf{94}} & \multirow{2}{*}{\textcolor{blue}{2}} \\ [2ex]\hline
				\multicolumn{1}{|c|}{\multirow{2}{*}{$ 315^0 $}} & \multirow{2}{*}{\textcolor{gray}{0}}  & \multirow{2}{*}{\textcolor{gray}{0}}  & \multirow{2}{*}{\textcolor{gray}{0}}  &\multirow{2}{*}{\textcolor{gray}{0}}   & \multirow{2}{*}{\textcolor{blue}{3}} & \multirow{2}{*}{\textcolor{gray}{0}}  &\multirow{2}{*}{\textcolor{gray}{0}}  & \multirow{2}{*}{\textbf{97}} \\[2ex] \hline
			\end{tabular}}
		\end{table}

		\section{Conclusion} \label{conc}
		
		This paper has proposed a novel DF method to be used in a drone surveillance system. The system comprises  a single channel receiver and a directional antenna array.  The receiver sequentially activates each antenna in the array, and measures the received power values. The power measurements  corresponding to each switching  cycle are fed to the proposed deep network. Then, it performs  DF by exploiting the sparsity property of the incoming drone signal, and the  gain variation property of the  directional antenna array. The paper has validated the proposed method experimentally.
		Also, it has been proven that a  phase synchronization mechanism, an antenna gain calibration mechanism, and the analytical model of  the antenna radiation pattern  are not essential for this single channel implementation. In future the scheme will be applied to portable, SDR-based
		prototype design \cite{MagDrone} and field test and experiment.
		%Hence, the proposed method is practical and low-complex. 

		% Below is an example of how to insert images. Delete the ``\vspace'' line,
		% uncomment the preceding line ``\centerline...'' and replace ``imageX.ps''
		% with a suitable PostScript file name.
		% -------------------------------------------------------------------------

		%\begin{figure*}[t] \vspace{0.3cm}
		%	\centering {\includegraphics[scale=0.9]{new10}} 
		%	\caption{System model.}	 
		%	\label{sm}
		%\end{figure*}

		%\begin{figure}[t]
		%	
		%	\begin{minipage}[b]{1.0\linewidth}
		%		\centering
		%		\centerline{\includegraphics[width=8.5cm]{image1}}
		%		%  \vspace{2.0cm}
		%		\centerline{(a) Result 1}\medskip
		%	\end{minipage}
		%	%
		%	\begin{minipage}[b]{.48\linewidth}
		%		\centering
		%		\centerline{\includegraphics[width=4.0cm]{image3}}
		%		%  \vspace{1.5cm}
		%		\centerline{(b) Results 3}\medskip
		%	\end{minipage}
		%	\hfill
		%	\begin{minipage}[b]{0.48\linewidth}
		%		\centering
		%		\centerline{\includegraphics[width=4.0cm]{image4}}
		%		%  \vspace{1.5cm}
		%		\centerline{(c) Result 4}\medskip
		%	\end{minipage}
		%	%
		%	\caption{Example of placing a figure with experimental results.}
		%	\label{fig:res}
		%	%
		%\end{figure}

		%\vfill
		%\pagebreak

		%
		%\vfill 
		%\pagebreak

\balance
\footnotesize {\bibliography{bibfile}}

\end{document}